\documentclass[pra,twocolumn,showpacs,nofootinbib,floatfix,superscriptaddress]{revtex4}

\usepackage{amsmath,amsfonts,amssymb,bm}
\usepackage{dcolumn}
\usepackage[final]{graphicx}

\begin{document}

\bibliographystyle{apsrev}

\title{Spin flip lifetimes in superconducting atom chips: BCS versus
Eliashberg theory}

\author{Ulrich Hohenester}\email{ulrich.hohenester@uni-graz.at}
\author{Asier Eiguren}
\affiliation{Institut f\"ur Physik,
  Karl--Franzens--Universit\"at Graz, Universit\"atsplatz 5,
  8010 Graz, Austria}

\author{Stefan Scheel}
\author{E. A. Hinds}
\affiliation{Quantum Optics and Laser Science, Blackett Laboratory,
Imperial College London, Prince Consort Road, London SW7 2AZ, United
Kingdom}

\date{\today}

\begin{abstract}
 We investigate theoretically the magnetic spin-flip transitions of
neutral atoms trapped near a superconducting slab. Our calculations
are based on a quantum-theoretical treatment of electromagnetic
radiation near dielectric and metallic bodies. Specific results are
given for rubidium atoms near a niobium superconductor.  At the low
frequencies typical of the atomic transitions, we find that BCS
theory greatly overestimates coherence effects, which are much less
pronounced when quasiparticle lifetime effects are included through
Eliashberg theory. At 4.2 K, the typical atomic spin lifetime is
found to be larger than a thousand seconds, even for
atom-superconductor distances of one micrometer. This constitutes a
large enhancement in comparison with normal metals.

\end{abstract}



\maketitle


\section{Introduction}

Over the last few years, enormous progress has been made in magnetic
trapping of ultracold neutral atoms near microstructured solid-state
surface, sometimes known as atom chips
\cite{hinds:99,folman:02,specialissue,fortagh:07}. The atoms can be
manipulated through variation of the magnetic confinement potential,
either by changing currents through gate wires mounted on the chip
or by modifying the strength of additional radio-frequency control
fields. These external, time-dependent parameters thus provide a
versatile method of atom manipulation, and make atom chips
attractive for various applications, including atom interferometry
\cite{haensel:01,hinds:01,andersson:02,wang:05,schumm:05,jo:07},
quantum gates
\cite{calarco:00,charron:06,treutlein:06,hohenester.pra:07} and
coherent atom transport \cite{paul:05}.  In addition, the atoms may
be used as a sensitive probe of the electromagnetic properties of
the surface in the neV (MHz) energy range.  For example they can
image surface currents in a normal metal \cite{wildermuth:05} or
vortices and flux noise in a type-II superconductor
\cite{scheel:07}.

On the other hand, the proximity of the ultracold atoms to the
solid-state structure, introduces additional decoherence channels,
which limit the performance of the atoms. Most importantly,
Johnson-Nyquist noise currents in the dielectric or metallic surface
arrangements produce magnetic-field fluctuations at the positions of
the atoms. Upon undergoing spin-flip transitions, the atoms become
more weakly trapped or are even lost from the microtrap
\cite{jones:03,lin:04}. Typically, the spin-flip transition
frequencies for magnetically trapped alkali atoms are in the sub-MHz
range, and the radiation-atom coupling is therefore strongly
enhanced by being in the near field regime
\cite{henkel:99,rekdal:04,scheel:05}. For atom-surface distances of
the order of one micrometer, the atom lifetime typically drops below
one second, which constitutes a serious limitation for atom chips.
It was shown in Ref.~\cite{scheel:05} that in order to reduce the
spin decoherence of atoms outside a metal in the normal state, one
should avoid materials whose skin depth at the spin-flip transition
frequency is comparable with the atom-surface distance. For typical
experimental designs using metals such as copper or gold, however,
the atom-surface distances are precisely in this range
\cite{jones:03,lin:04}.

Superconductors could reduce the magnetic noise level significantly
and thereby boost the spin flip lifetimes by many orders of
magnitude. Indeed, superconducting atom chips have already been
fabricated and tested \cite{nirrengarten:06,shimizu:07} with the aim
of realizing controllable composite quantum systems. Previous
estimates of the lifetime enhancement relative to a normal metal
surface have given factors of tens \cite{scheel:05} or millions
\cite{skagerstam:06}, depending on the theoretical approach. Scheel
{\em et al.}\/ \cite{scheel:05} considered the energy dissipation in
the superconducting state resulting from the modified quasiparticle
dispersion, whereas Skagerstam {\em et al.}\/ \cite{skagerstam:06}
considered the screening of the current fluctuations by the
superconductor. The two approaches are difficult to compare, since
they ignore the strong modification of either the imaginary part
\cite{scheel:05} or the real part \cite{skagerstam:06} of the
optical conductivity in the superconducting state. The question of
how to describe the problem properly led to some dispute
\cite{scheel.comment:06,skagerstam.reply:06}.

In this paper, we resolve the dispute and present a scheme for the
proper description of magnetic spin-flip rates in atoms on a
superconducting atom chip. Our analysis is based on three
descriptions of superconductivity. We start with the two-fluid model
\cite{skagerstam:06}, and then progress via the
Bardeen--Cooper--Schrieffer (BCS) theory \cite{bardeen:57} to a more
elaborate framework, the Eliashberg theory \cite{eliashberg:60},
which we find is needed for a proper description of this problem.
For typical spin flip frequencies on a chip
($1\,\mbox{kHz}$--$10\,\mbox{MHz}$), we point out that the BCS
theory significantly overestimates the optical conductivity and
hence gives too high a value for the spin flip rate. A realistic
calculation of the conductivity \cite{marsiglio:02} requires further
elaboration, in the framework of the Eliashberg theory, to include
lifetime effects of the quasiparticles due to phonon scattering.
This results in a reliable estimate of the spin flip rate, which
ends up not far from the two-fluid result of Skagerstam {\em et al.}
\cite{skagerstam:06}. We conclude that superconducting surfaces can
be used to achieve low spin flip rates in an atom chip, with
lifetimes exceeding a thousand seconds for Rb atoms at $1\,\mu$m
from a Nb surface at $4.2\,$K.

We have organized our paper as follows. In Sec.~II we present the
three models for the description of superconductors. Although there
is already a vast literature on superconductivity, including many
textbooks \cite{rickayzen:65,tinkham:75,mahan:81}, we give next a
brief account of these approaches, mainly to make the paper as
self-contained as possible. We discuss the basic assumptions of the
two-fluid model, the appearance and shortcoming of the coherence
peak in BCS theory, and the description of quasiparticle damping and
formation within the framework of Eliashberg theory. In Sec.~III we
present our results for the lifetime of an atom placed in the
vicinity of a semi-infinite niobium sample. We compare the different
approaches and discuss their respective advantages and
disadvantages. Finally, we summarize our results in Sec.~IV.


\section{Theory of superconductivity}

\subsection{Two-fluid model}

The first successful attempt to account for the electromagnetic
properties of superconductors was due to F. and H.~London
\cite{london:35}. They devised a phenomenological two-fluid model
that was able to explain many of the phenomena observed in
superconductors.

Within this model one assumes that there are two types of charge
carrier, superconducting and normal, which react differently to
external electromagnetic fields. We write $n_n(T)$ and $n_s(T)$ to
denote the electron number densities in the normal and
superconducting states at temperature $T$, with $n_n(T)+n_s(T)=n_0$
assumed to be constant. Although it does not become obvious from the
two-fluid model itself, the superconducting carriers have to be
associated with Cooper pairs. At temperatures above the
superconductor transition temperature $T_c$, only normal carriers
are present and $n_n(T\!\!>\!T_c)=n_0$, while at zero temperature
all carriers are in the superconducting state, $n_s(0)=n_0$.

For the normal electrons, the response to a sufficiently weak
external electric field $\mathbf{E}$ is given by Ohm's law
$\mathbf{j}_n=\sigma_n\mathbf{E}$, with $\mathbf{j}_n$ being the
current density of the normal electrons and $\sigma_n$ the
normal-state conductivity. For the superconducting current
$\mathbf{j}_s$, the London brothers introduced a new relation
\begin{equation}\label{eq:london}
  \Lambda\frac{\partial\mathbf{j}_s}{\partial t}=\mathbf{E}\,,
\end{equation}
where $\Lambda$ is a constant whose value varies for different
superconducting materials. This describes the dynamics of carriers
that are accelerated freely in an electric field. For a
superconductor made up of free electrons (or indeed of free Cooper
pairs), the value of $\Lambda$ would be $m/(n_se^2)$, where $m$ and
$e$ are the single electron mass and charge. In fact this also
provides a useful estimate for real superconductors.
Later in the paper we will re-write this relation in terms of the
plasma frequency $\omega_p$, as $\Lambda\simeq
1/(\varepsilon_0\omega_p^2)$. As a consequence of the London
equation, \eqref{eq:london}, a static magnetic field can only
penetrate into a superconductor by a distance of order
$\lambda_L=(\Lambda /\mu_0)^\frac 12$ \cite{rickayzen:65}. For this
reason $\lambda_L$ is called the penetration depth or London length.

Consider an electric field oscillating as $\exp(-i\omega t)$. The
response of the superconductor is given by
\begin{equation}
  \mathbf{j}=\mathbf{j}_n+\mathbf{j}_s
  =\left(\sigma_n+\frac i{\omega\Lambda}\right)\mathbf{E}\,.
\end{equation}
Here, the expression in parentheses
\begin{equation}\label{eq:opcond}
  \sigma(\omega)\equiv \sigma'(\omega)+i\sigma''(\omega)=
  \frac{1}{\omega\mu_0}
  \left(\frac 2{\delta^2}+\frac i{\lambda_L^2}\right)\,,
\end{equation}
is known as the optical conductivity, though in this paper we will
be using it at radio frequencies. We have introduced the skin depth
$\delta=(2/\mu_0\omega\sigma_n)^\frac 12$ associated with the normal
charge density.

For the two-fluid model, Eq.~\eqref{eq:opcond} can be further
simplified by noting that the two contributions vary with
temperature only through the normal and superconducting charge
densities. Thus, with $\sigma_0$ being the conductivity in the
normal state and $\Lambda_0$ the $\Lambda$-parameter at zero
temperature, we have
\begin{equation}\label{eq:opcond2}
  \sigma(\omega)\cong
\sigma_0 \,\frac{n_n(T)}{n_0}+
  \frac{i}{\omega\Lambda_0}\,\left(1-\frac{n_n(T)}{n_0}\right)\,.
\end{equation}
For $T<T_c$, a suitable form for the temperature dependence of the
normal density is provided by the Gorter-Casimir expression
$n_n(T)=(T/T_c)^4\, n_0$ \cite{gorter:34}.

\subsection{Bardeen--Cooper--Schrieffer (BCS) theory}
\label{sec:bcs}

Despite its success, the London theory has a number of shortcomings.
First, it is phenomenological and not based on a microscopic model.
Second, its predictions cannot account for all experimental
observations. A relevant example here is its inability to account
for the so-called {\em coherence peak},\/ that was first observed in
NMR by Hebel and Slichter \cite{hebel:59}. This peak is most
pronounced at low frequencies and is thus of importance for the
analysis of spin decoherence in superconducting atom chips. In order
to understand its origin we introduce the theory of  Bardeen,
Cooper, and Schrieffer (BCS) \cite{bardeen:57}.

\subsubsection{BCS ground state}

BCS theory is based on Fr\"ohlich's observation \cite{froehlich:50}
that electrons close to the Fermi energy $\epsilon_F$ can attract
each other through the exchange of virtual phonons, and Cooper's
demonstration \cite{cooper:56} that due to this interaction the
Fermi sea is unstable against the formation of a certain kind of
quasi-bound pair. The attractive electron-electron interaction is
usually described by the pairing Hamiltonian \cite{rickayzen:65}
\begin{equation}\label{eq:hamp}
  H_p=\sum_{\mathbf{k}\sigma}\xi_{\mathbf{k}}^{\phantom{\dagger}}\, 
  c_{\mathbf{k}\sigma}^\dagger
  c_{\mathbf{k}\sigma}^{\phantom\dagger}-V{\sum_{\mathbf{k},\mathbf{k}'}}'
  c_{\mathbf{k}\uparrow}^\dagger c_{-\mathbf{k}\downarrow}^\dagger
  c_{-\mathbf{k}'\downarrow}^{\phantom\dagger}
  c_{\mathbf{k}'\uparrow}^{\phantom\dagger}
\,.
\end{equation}
Here $c_{\mathbf{k}\sigma}^\dagger$ is the field operator for the
creation of an electron with wavevector $\mathbf{k}$ and spin
orientation $\sigma$,
$\xi_{\mathbf{k}}=\epsilon_{\mathbf{k}}-\epsilon_F$ is
the single-electron energy $\epsilon_{\mathbf{k}\sigma}$ measured
with respect to $\epsilon_F$, and $V$ is the strength of the
attractive phonon-mediated electron-electron interaction. The prime
on the sum indicates that this interaction has to be considered only
for electrons with energy smaller than the Debye energy
$\hbar\omega_D$.

As result of this coupling, electrons are promoted from states below
the Fermi energy to states above to form Cooper pairs. This process
comes to a halt when the increase in kinetic energy is no longer
compensated by the reduction in potential energy from the pairing.
To model the phase transition associated with the formation of
Cooper pairs, one assumes that the interaction operator $c_{-\bm
k\downarrow}^{\phantom\dagger}c_{\mathbf{k}\uparrow}^{\phantom\dagger}$
is practically a $c$-number $b_{\mathbf{k}}^0$, with small
fluctuations about this value. One then formally writes all pairs of
operators in the form
$c_{-\mathbf{k}\downarrow}^{\phantom\dagger}c_{\bm
k\uparrow}^{\phantom\dagger} =b_{\mathbf{k}}^0+(c_{-\bm
k\downarrow}^{\phantom\dagger}c_{\mathbf{k}\uparrow}^{\phantom\dagger}
-b_{\mathbf{k}}^0)$ and neglects the terms bilinear in the
parenthetical quantities. The resulting mean-field Hamiltonian can
be diagonalized through a Bogoliubov transformation
\begin{displaymath}
  c_{\mathbf{k}\uparrow}^{\phantom\dagger}=
  u_{\mathbf{k}}\gamma_{\mathbf{k}0}^{\phantom\dagger}+
  v_{\mathbf{k}}^*\gamma_{\mathbf{k}1}^\dagger\,,\quad
  c_{-\mathbf{k}\downarrow}^\dagger =
  -v_{\mathbf{k}}\gamma_{\mathbf{k}0}^{\phantom\dagger}+
  u_{\mathbf{k}}^*\gamma_{\mathbf{k}1}^\dagger\,,
\end{displaymath}
where $\gamma_{\mathbf{k}0}^\dagger$ and
$\gamma_{\mathbf{k}1}^\dagger$ create Fermionic quasiparticles that
are linear superpositions of the bare electron states, and the
coefficients $u_{\mathbf{k}}$ and $v_{\mathbf{k}}$ are chosen to
diagonalize the Hamiltonian,
\begin{equation}\label{eq:hambcs}
  H_{\rm BCS}={\sum_{\mathbf{k}}}' E_{\mathbf{k}}\left(
  \gamma_{\mathbf{k}0}^\dagger \gamma_{\mathbf{k}0}^{\phantom\dagger}+
  \gamma_{\mathbf{k}1}^\dagger \gamma_{\mathbf{k}1}^{\phantom\dagger}\right)+
  \mbox{const}\,.
\end{equation}
Here, $E_{\mathbf{k}}=(\xi_{\mathbf{k}}^2+\Delta^2)^\frac 12$ are the
new quasiparticle excitation energies in the superconducting state,
and $\Delta=V{\sum_{\mathbf{k}}}'b_{\mathbf{k}}^0$ is the order
parameter or gap parameter. $\Delta$ has to be determined from the
numbers $b_{\mathbf{k}}^0$ which are the thermal and quantum averages of
\begin{equation}\label{eq:gapfun}
  b_{\mathbf{k}}^0=\mbox{tr}\left(e^{-\beta H_{\rm BCS}}
  c_{-\mathbf{k}\downarrow}^{\phantom\dagger}
  c_{\mathbf{k}\uparrow}^{\phantom\dagger}
  \right)/\mbox{tr}\,e^{-\beta H_{\rm BCS}}\,,
\end{equation}
where $\beta\equiv 1/(k_B T)$. Equation~\eqref{eq:gapfun} is a
self-consistency relation, since the values of $b_{\mathbf{k}}^0$
are hidden within $H_{\mbox{\tiny BCS}}$ through its dependence on
the quasiparticle energies $E_{\mathbf{k}}$.
Thermally excited quasiparticles with energy $E_{\mathbf{k}}$
restrict the phase space available for forming Cooper pairs and
thereby reduce the gap parameter $\Delta$.

\subsubsection{Coherence peak}
\label{subsec:coherence}

The density of these quasiparticle states at energy $E$ is given by
\cite{rickayzen:65}
\begin{equation}\label{eq:qpdos}
  \rho(E) =
  \left\{
  \begin{array}{ll}
  N(\epsilon_F)\displaystyle\frac{E}{\sqrt{E^2-\Delta^2}} \,,& E\ge \Delta \,,\\
  0 \,, & E<\Delta \,.
  \end{array}
  \right.
\end{equation}
At zero temperature, no quasiparticles are excited and therefore the
only way to deposit energy in the superconductor is to break up
Cooper pairs. Consequently the real part $\sigma'$ of the $T=0$
conductivity is strictly zero for electric field frequencies below
$2\Delta/\hbar$. At non-zero temperatures however, many
quasiparticles may be excited just above the gap because the density
of states is so high there---indeed $\rho(E)$ diverges in
Eq.~\eqref{eq:qpdos} at $E=\Delta$. This opens up a mechanism for
dissipation at low frequency. The corresponding $\sigma'$ involves
the density of quasiparticles, which is proportional to $\rho(E)$,
and the density of final states for absorption of a photon at
frequency $\omega$, which is proportional to $\rho(E+\hbar \omega)$.
Integration over $E$ produces a logarithmically divergent
conductivity $\sigma'(\omega)\sim \sigma_0 \ln(2\Delta/\omega)$.
This enhancement, which was first observed in nuclear magnetic
resonance \cite{hebel:59}, is known as the Hebel-Slichter or
coherence peak. This reasoning is supported by Mattis and Bardeen's
expression for the optical conductivity \cite{mattis:58}, which was
computed with the random-phase approximation and in the dirty limit,
where scattering by impurities reduces the coherence length to less
than the magnetic-field penetration length $\lambda_L$. This gives
the same logarithmic divergence of $\sigma'(\omega)$ at low
frequency \cite{tinkham:75,klein:84}. At zero frequency, we note
that $\sigma'$ has another singularity of $\delta$ type, associated
with the dc response of the superfluid.

For the sub-MHz spin-flip transitions of magnetically trapped
ultracold atoms, the BCS theory thus predicts a strong modification
of the optical conductivity in comparison to the
frequency-independent value of Eq.~\eqref{eq:opcond2} given by the
two-fluid model: $\sigma'=\sigma_0 (n_n/n_0)$.

\subsection{Eliashberg theory}

While the BCS theory incorporates the mixing of free electron states
through their coupling to virtual phonons, it does not include the
dissipative effects associated with the emission and absorption of
real phonons. This broadens the quasiparticle states and softens the
divergence of the conductivity at low frequency so that it is much
less dramatic.

Phonon scattering converts the electron wavevector {\bf k} to
wavevectors {\bf k$'$} at a rate $1/\tau_{\bf k}$, given by Fermi's
Golden Rule as
\begin{eqnarray}\label{eq:fermigoldenrule}
\frac{1}{\tau_{\bf k}}\cong \frac{2\pi}{\hbar} \sum_{\bf k'} \left |
g^{\lambda}_{\bf k,k'} \right |^2 \delta(\epsilon_{\bf
k}-\epsilon_{\bf k'}-\hbar\omega^{\lambda}_{\bf q}) \nonumber\\
\times \left[2\bar n_{\rm th}(\hbar\omega^{\lambda}_{\bf
q})+1\right]\,,
\end{eqnarray}
where $\hbar\omega^{\lambda}_{\bf q}$ is the energy of a phonon in
mode $\lambda$  with wavevector ${\bf q=k-k'}$, $\bar n_{\rm th}$ is
the number of thermal phonons in the mode and $g^{\lambda}_{\bf
k,k'}$ is the off-diagonal matrix element of the the electron-phonon
interaction Hamiltonian. Since electron energies are typically two
orders of magnitude larger than the Debye energy, the phonon
energies entering the Dirac delta function in
Eq.~\eqref{eq:fermigoldenrule} can be safely neglected. This
approximation leads one to define the dimensionless quantity
\begin{eqnarray}\label{eq:eliashberg}
\alpha^2 F_{\bf k}(\omega) = \sum_{\bf k', \lambda} \left | g^{\lambda}_{\bf
k,k'} \right |^2
\delta(\epsilon_{\bf k'}-\epsilon_F) \delta(\omega-\omega^{\lambda}_{\bf q})\,,
\end{eqnarray}
known in literature as the Eliashberg function \cite{mahan:81}.
Thus, the electron scattering rate at low temperatures can be
conveniently written as
\begin{eqnarray}\label{lifetime}
1/\tau_{\bf k}\approx \frac{2\pi}{\hbar} \int_{0}^{\omega_D} d
\omega \ \alpha^2 F_{\bf k}(\omega) \left[2
n(\hbar\omega)+1\right]\,.
\end{eqnarray}
In this expression, the Eliashberg function encapsulates all the
relevant information about the electron-phonon coupling and the Fermi
surface. The complex self energy $\Sigma$ resulting from this coupling
gives both the scattering rate that we have just discussed, through
$\hbar/\tau_{\bf k}=2\, \Im m[\Sigma(\epsilon_{\bf k})]$, and the
energy shift $\Re e[\Sigma(\epsilon_{\bf k})]$ of the electron.

There are two kinds of self-energy function in the description of a
superconductor, usually labeled normal and anomalous. The normal
component has the same meaning as in an ordinary metal, whereas the
anomalous one is directly related to the opening of the gap due to
the formation of Cooper pairs. These are closely related because the
scattering rate and distortion of the electron bands due to the
electron-phonon coupling depend strongly on the superconducting gap,
and vice versa. This interdependence is accounted for by the
so-called Eliashberg equations, which must be solved
self-consistently \cite{mahan:81}.

A powerful numerical implementation for the solution of the
Eliashberg equations has been developed by Carbotte, Marsiglio, and
coworkers \cite{marsiglio:88,carbotte:90,nicol:92,marsiglio:02},
where one first computes the electron Green function in Matsubara
space and then performs an analytic continuation by means of an
iterative procedure. The real-frequency-axis Green functions can be
used finally to compute the optical conductivity
\cite{nam:67,nicol:92,marsiglio:02}, including not only
electron-phonon interactions, described above, but also the effects
of elastic impurity scattering.

\subsection{Impurity effects}\label{sec:impurity}

We conclude this section by briefly addressing effects due to
elastic impurity scattering. In conventional superconductors
impurities are deemed to be innocuous as a result of Anderson's
argument \cite{anderson:59,balatsky:06}, which goes as follows. In
the normal state, the electrons can be described by wavefunctions
$\phi_{n\uparrow}({\bf r})$ and $\phi_{n\downarrow}({\bf r})$, where
$\phi_n$ is supposed to include the effects of impurity scattering.
The quantum number $n$ replaces the wavenumber ${\bf k}$ of the pure
metal. In the pure superconductor, the Cooper pair is composed of
the states $({\bf k},\uparrow)$ and $(-{\bf k},\downarrow)$.
Anderson pointed out that the second of these states is the first
with momentum and current reversed in time. In an impure
superconductor the main contribution to the pairing should be also
between the time-reversed states $\phi_{n\uparrow}({\bf r})$ and
$\phi_{n\downarrow}^*({\bf r})$. The pairing Hamiltonian
\eqref{eq:hamp} can thus be expressed in terms of the new operators
$c_{n\sigma}^{\phantom\dagger}$ and $c_{n\sigma}^\dagger$, where the
interaction matrix element between two states becomes
\begin{equation}
  V_{nn'}=V\sum_{{\bf k},{\bf k}'}\left|\langle n|{\bf k}\rangle\right|^2
  \left|\langle n'|{\bf k}'\rangle\right|^2=V\,.
\end{equation}
Owing to the completeness relation of the states involved, the
pairing Hamiltonian is not modified in the new basis $\phi_n$. For
this reason the superconductor properties such as, e.g., transition
temperature, gap parameter, or quasiparticle density of states, are
not significantly changed by the presence of impurities.

The argument above applies not only to BCS but also to Eliashberg
theory as long as the Eliashberg function $\alpha^2F_{\bf
k}(\omega)$ has little dependence on the direction of $\bf k$. This
is indeed the case for the conventional $s$-wave superconductors we
are considering. Moreover, any small anisotropy is randomized by the
impurity scattering, so it suffices in this work to consider the
average over all directions $\alpha^2F(\omega)=\left<\alpha^2F_{\bf
k}(\omega)\right>$.

Although impurities do not affect the pairing Hamiltonian, the
scattering from impurities at rate $\gamma$ plays an important role
in the electron transport because the normal conductivity $\sigma_0$
is approximately inversely proportional to $\gamma$. In the
two-fluid model and in BCS theory, $\sigma'(\omega)$ increases in
direct proportion to $\sigma_0$ as the scattering rate is reduced.
In Eliashberg theory however, the situation is complicated by the
presence of the inelastic phonon scattering, which tends to reduce
the conductivity through the broadening of the density of
quasiparticle states. As $\gamma$ is reduced, this effect becomes
relatively more important, causing $\sigma'(\omega)$ to increase
more slowly than $\sigma_0$. In the calculations that follow, we
will allow $\gamma$ to be a variable in the optical conductivity
\cite{nicol:92,marsiglio:02} so that we can explore this effect. We
will find that this provides a connection between the two-fluid and
BCS results as well as allowing us to make contact with real
materials.

\section{Results for the atom trapping lifetime}

\begin{figure}
\centerline{\includegraphics[width=0.9\columnwidth]{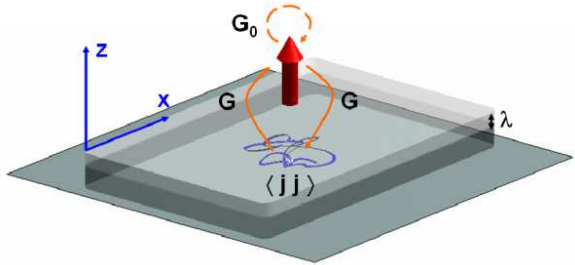}}
\caption{(color online) Schematic geometrical setup. A plane
metallic or superconducting slab lies parallel to the $(x,y)$ plane.
The atom with magnetic moment $\bm\mu$ indicted by the arrow, is
located in vacuum at a distance $z$ from the surface. The atom
suffers spontaneous or thermally stimulated magnetic spin-flip
transitions, as indicated by $\bm G_0$ and $\bm G$, thereby becoming
more weakly trapped and eventually lost. Johnson current noise
$\langle \mathbf{j} \mathbf{j} \rangle$ within the penetration depth
$\lambda$ contributes to magnetic-field fluctuations at the position
of the atom. }\label{fig:schematics}
\end{figure}

We turn now to the spin flip rate for an atom located in vacuum near
a superconducting slab, as illustrated in Fig.~\ref{fig:schematics}.
Following Refs.~\cite{rekdal:04,scheel:05,skagerstam:06}, we
consider a ground-state alkali atom, magnetically trapped in a
weak-field-seeking Zeeman sub-level. The noise in the magnetic
field, due both to vacuum fluctuations and to thermal currents in
the surface, induces transitions between the levels, making the
atomic spin change direction (spin flip) and ultimately causing the
atom to be lost from the microtrap.

As briefly outlined in Appendix \ref{app:lifetime}, the spin-flip
lifetime of an atom at position $\bm r_A$ is directly related to the
imaginary part of the dyadic Green tensor
$\bm{G}(\mathbf{r}_A,\mathbf{r}_A,\omega)$ of Maxwell's theory. The
usual, free-space spontaneous emission rate is determined by the
vacuum contribution $\bm{G}_0$. For a typical transition frequency
of $f_A=\omega_A/(2\pi)=500$ kHz, corresponding to an energy of
approximately 2 neV, this natural lifetime at zero temperature is
$\tau_0\approx 2\times 10^{25}$ seconds \cite{scheel:05}, which can
safely be considered infinite. The dominant contribution to the
lifetime reduction comes from the magnetic-field fluctuations
induced by the Johnson-Nyquist noise in the dielectric body. As
shown in Fig.~\ref{fig:schematics} and discussed in the Appendix,
the current noise translates through the Green tensors to a
magnetic-field fluctuation at the position of the atom.

For a thick superconducting slab described by an optical
conductivity in the limit $\sigma''(\omega)\gg\sigma'(\omega)$ and
in the near field regime $\lambda_L\ll z\ll 2\pi/k$ one calculates,
using the results of Ref.~\cite{skagerstam:06}, a spin-flip rate of
\begin{equation}\label{eq:lifetime}
  \Gamma\equiv\frac{1}{\tau_A}\approx\Gamma_0\left(\bar n_{\rm th}+1\right)
  \left[ 1+ \frac{27}{64(\omega\mu_0)^{1/2}k^3 z^4}
  \frac{\sigma'}{(\sigma'')^{3/2}}
  \right]\,.
\end{equation}
Here $\tau_A$ is the spin flip lifetime, $\Gamma_0$ is the
free-space decay rate, $\bar n_{\rm th}$ is the mean thermal photon
number at the transition frequency $\omega_A$, $k=\omega_A/c$, and
$z$ is the atom-superconductor distance. In the following sections
we investigate the consequences for this rate of using the
expressions for the optical conductivity
$\sigma(\omega)=\sigma'(\omega)+i\sigma''(\omega)$ obtained from a
two-fluid description, from BCS theory and from Eliashberg theory.

\subsection{Two-fluid model}

To estimate the order of magnitude of these parameters, we first
consider the simple two-fluid model. Using the expression
(\ref{eq:opcond}) for the optical conductivity,
Eq.~(\ref{eq:lifetime}) reduces to the expression
\cite{skagerstam:06}
\begin{equation}
  \Gamma\equiv\frac{1}{\tau_A}\approx\Gamma_0\left(\bar n_{\rm th}+1\right)
  \left[1+2\left(\frac 34\right)^3
  \frac 1{k^3\delta^2}\frac{\lambda_L^3}{z^4}\right]\,.
\end{equation}
We take Nb as a representative superconducting material throughout.
Table~\ref{table:sigma} shows a few values reported in the
literature for the conductivity $\sigma_0$ of the normal state. We
note that the ultra-pure niobium sample of Ref.~\cite{casalbuoni:05}
has a hundred times higher conductivity than the films of
Refs.~\cite{perkowitz:85,pronin:98}. Through the simple Drude model
\cite{ashcroft:76}
\begin{equation}\label{eq:drude}
  \sigma_0=\varepsilon_0\omega_p^2\tau\,,
\end{equation}
we can relate $\sigma_0$ to an electron lifetime $\tau=1/\gamma$ due
to elastic scattering at impurities or defects. $\hbar\omega_p$ is
the bulk plasmon energy, which we set equal to 10 eV
\cite{ashcroft:76,perkowitz:85}. The corresponding $\tau$ values are
given in the last column of Table~\ref{table:sigma}. With an atomic
transition frequency of 500 kHz, we obtain for the ultrapure sample
a normal-state skin depth of
$\delta_0=\sqrt{2/(\mu_0\omega\sigma_0)}\approx 16$ $\mu$m and a
value approximately ten times larger for the films.

\begin{table}
\caption{Normal-state conductivity $\sigma_0$ measured on several
different samples of niobium. The approximate scattering times
$\tau$ are obtained from the Drude model \eqref{eq:drude}. The
corresponding plasma frequency is $\omega_p\approx 10\,
\mbox{eV}/\hbar\simeq 1.5\times
10^{16}\mbox{s}^{-1}$.}\label{table:sigma}
\begin{ruledtabular}
\begin{tabular}{lcc}
Reference & $\sigma_0$ ($\mu\Omega^{-1}{\rm cm}^{-1}$) & $\tau$ (fs) \\
\colrule
Perkowitz \textit{et al.}~\cite{perkowitz:85} & 0.2 & 10 \\
Pronin \textit{et al.}~\cite{pronin:98} & 0.25 & 13 \\
Klein \textit{et al.}~\cite{klein:84} & 0.85 & 43 \\
Casalbuoni \textit{et al.}~\cite{casalbuoni:05} & 20 & 1000\\
\end{tabular}
\end{ruledtabular}
\end{table}

A rough estimate for the penetration depth of the superconductor at
zero temperature is given by
\begin{equation}\label{eq:lambdaapprox}
  \lambda_L=\left(\frac{\Lambda}{\mu_0}\right)^{\frac 12}\approx
  \left(\frac{1}{\mu_0\,\varepsilon_0\,\omega_p^2}\right)^{\frac 12}=
  \frac c{\omega_p}\approx 20\,\mbox{nm}\,,
\end{equation}
where we have assumed that all electrons move freely. This simple
estimate is comparable to the BCS value of 35 nm \cite{miller:59},
and to the experimental values of 46 nm for the ultrapure sample
\cite{casalbuoni:05} and 90 nm for the niobium film in
\cite{pronin:98}.

\subsection{BCS versus Eliashberg theory}

\begin{figure}
\centerline{\includegraphics[width=0.9\columnwidth]{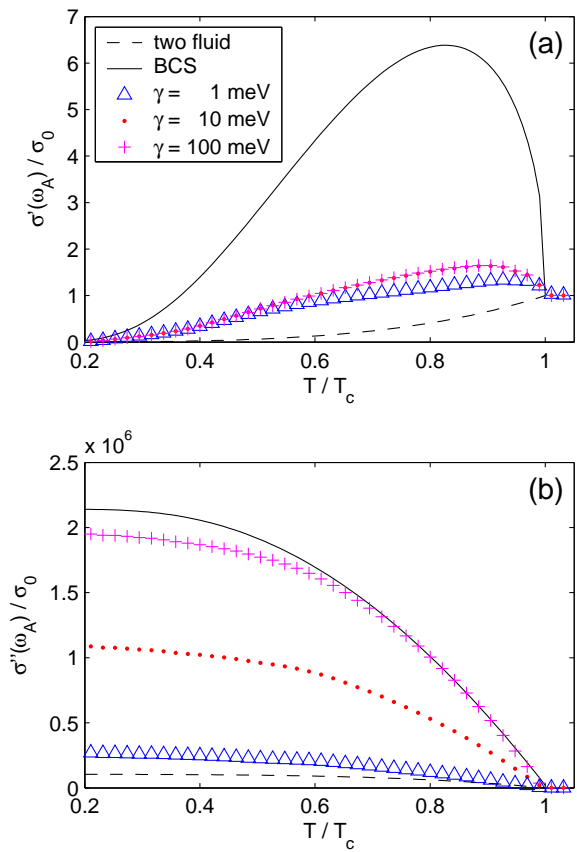}}
\caption{(color online) Temperature dependence of (a) real part
$\sigma'(\omega_A)$ and (b) imaginary part $\sigma''(\omega_A)$ of
the optical conductivity, normalised to the normal state
conductivity $\sigma_0$. $\omega_A=2\pi\times 500\mbox{kHz}$ is the
atomic spin-flip frequency, $T_c=9.2$ K is the superconductor
transition temperature. The different lines correspond to the
results for the two-fluid model (dashed line), using $\delta_0=16$
$\mu$m and $\lambda_L(0)=35$ nm, BCS theory (solid line), and
Eliashberg theory (symbols) for three elastic impurity scattering
rates $\gamma$.} \label{fig:sigma500kHz}.
\end{figure}

Now we discuss how the two-fluid estimates are modified within the
framework of BCS and Eliashberg theories. For the BCS theory of
niobium we use a zero-temperature gap parameter of $\Delta=1.4$ meV,
corresponding to a transition temperature of $T_c=9.2$ K, and a
Debye temperature of $\hbar\omega_D/k_B=275$ K, and we compute the
optical conductivity by means of the Mattis--Bardeen formulas in the
dirty limit \cite{mattis:58}.

For the implementation of the Eliashberg equations, we have
considered an $\alpha^2F(\omega)$ function calculated using linear
response theory \cite{baroni:01} and norm-conserving
pseudopotentials. The electron-phonon matrix elements were
calculated on a 32$^3$ wavevector grid for both electrons and
phonons. Our result (not shown) is similar to that presented in
Ref.~\cite{savrasov:96}, though with spectral features that are less
pronounced, in better agreement with the data of tunneling
experiments. As far as the calculated atomic spin flip rates are
concerned, we do not find any significant difference between these
two $\alpha^2F(\omega)$ functions.

Figure \ref{fig:sigma500kHz} shows results for the (a) real and (b)
imaginary part of the optical conductivity versus temperature $T$.
The solid lines show the results from BCS theory in the dirty limit,
the dashed lines are for the two-fluid model and the symbol series
are for Eliashberg theory with various values of the elastic
impurity scattering rate $\gamma$. In Fig. \ref{fig:sigma500kHz}(a),
the $\sigma'(\omega_A)$ obtained from the two-fluid model decreases
monotonically with temperature because of the decrease in the normal
density $n_n(T)$, whereas the BCS and Eliashberg curves show an
enhancement of $\sigma'(\omega_A)$ at temperatures immediately below
the transition temperature $T_c$. This is due to the coherence peak,
which forms as a consequence of the modified quasiparticle
dispersion in the superconducting state. The peak is most pronounced
within the BCS framework in the dirty limit. As we move away from
the dirty limit towards a clean superconductor, using the Eliashberg
theory with decreasing rates $\gamma$, we observe that the peak
gradually disappears, in agreement with \cite{marsiglio:94}. Thus
the Eliashberg theory interpolates between the two extreme cases of
the two-fluid model and the dirty limit of BCS theory by varying the
chosen value of $\gamma$.

In Fig.~\ref{fig:sigma500kHz}(b) we show the imaginary part of the
optical conductivity. Again, the BCS result is according to the
theory of Mattis and Bardeen \cite{mattis:58} for a dirty
superconductor.  Here too, we see that the Eliashberg theory with
variable $\gamma$ provides a link between the two-fluid and BCS
extremes. In the low frequency limit, the BCS result takes the
analytical form
\begin{equation}\label{eq:sigma''}
  \sigma_{\rm BCS}''(\omega)=\sigma_0\,\frac{\pi\Delta}{\hbar\omega}\,
  \tanh\frac{\Delta}{2k_BT}\,,
\end{equation}
where $\Delta$ is the temperature-dependent gap parameter.  We note
that this $1/\omega$ dependence of $\sigma''$ is the same in all
three models. This is a consequence of the Kramers-Kronig relations
together with the fact that $\sigma'$ has a $\delta$ singularity at
$\omega=0$ associated with the response of the superfluid to a dc
field.

We are using here a theory in which the material responds locally to
a field. Although this is not strictly so, nonlocality can be
incorporated empirically into the theory of the superconductor
through a modified penetration depth \cite{skagerstam:06}. The
effect of nonlocality on the atom-surface response is negligible
since the penetration depth is small compared with the
atom-superconductor distance.

\begin{figure}
\centerline{\includegraphics[width=0.9\columnwidth]{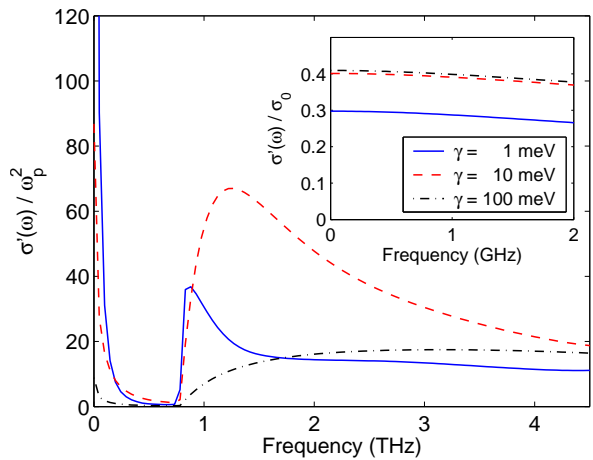}}
\caption{(color online) $\sigma'(\omega)$ at 4.2 K as a function of
frequency for three elastic scattering rates $\gamma$, as computed
within the framework of Eliashberg theory. The peak at zero
frequency is attributed to the condensate, and the peak at 1 THz to
the breaking of Cooper pairs. In the inset we show that the
condensate peak saturates at low frequencies.}\label{fig:sigma4K}
\end{figure}

As discussed in Sec.\ref{subsec:coherence}, the BCS coherence peak
illustrated by the solid line in Fig.~\ref{fig:sigma500kHz}(a)
increases with decreasing frequency, diverging as $\omega\to 0$.
This behaviour is greatly suppressed when inelastic phonon
scattering is taken into account using Eliashberg theory (see also
the discussion in Sec.~\ref{sec:impurity}), as plotted in
Fig.~\ref{fig:sigma4K}. This figure shows the real part of the
optical conductivity $\sigma'(\omega)$ at 4.2 K with three values of
$\gamma$, spanning the $10-1000\,$ps range of scattering times given
in Table~\ref{table:sigma}. The peak in Figure~\ref{fig:sigma4K} at
1 THz is the conductivity associated with the breakup of Cooper
pairs. The lower-frequency peak, which is the one of relevance here,
no longer diverges at low frequency but reaches a constant value,
shown inset in the figure for frequencies below 2~GHz. Here, as in
Fig.~\ref{fig:sigma500kHz}, the value of $\sigma'$ is normalised to
the normal state conductivity to remove most of the dependence on
$\gamma$. Since the atomic spin flip frequency is bound to be in
this low frequency range, the Eliashberg results shown in
Fig.~\ref{fig:sigma500kHz} apply to all cases of experimental
interest. We recall that the $\sigma'$ of the two fluid model in Eq.
\eqref{eq:opcond} is also frequency independent.

\begin{figure}
\centerline{\includegraphics[width=0.9\columnwidth]{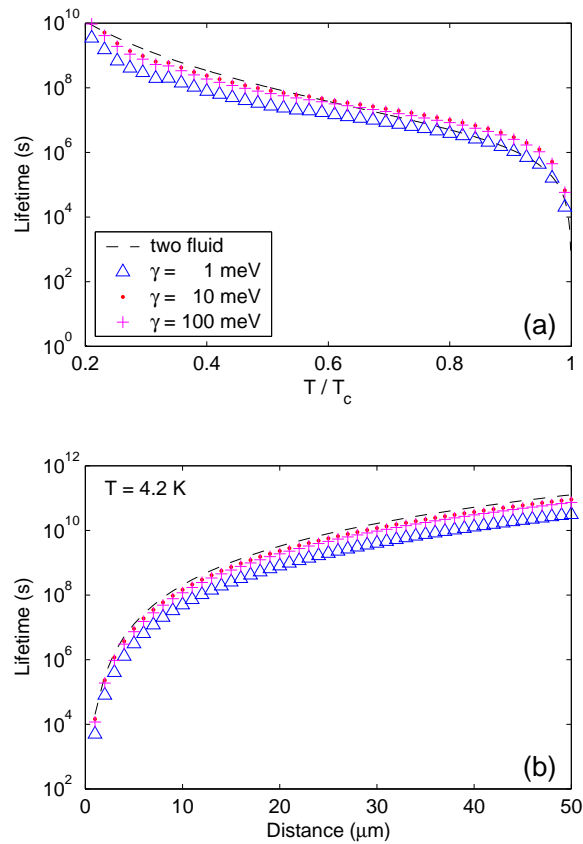}}
\caption{(color online) Spin-flip lifetime $\tau_A$ of a trapped
atom near a superconducting slab as a function of (a) temperature
(for 10 $\mu$m atom-surface distance), and (b) distance (at 4.2~K).
The smallest distance in (b) is 1 $\mu$m (with $\tau_A\approx 5000$
s for $\gamma=1$ meV). For the calculation of $\tau_A$ we use
Eq.~\eqref{eq:lifetime} and the optical conductivity computed within
Eliashberg theory for different elastic scattering rates $\gamma$.
The atomic transition frequency is fixed at 500~kHz throughout. The
dashed lines correspond to calculations performed with the two-fluid
model and the parameters given in
Ref.~\cite{skagerstam:06} (London length $\lambda_L=35$ nm).}%
\label{fig:lifetime}
\end{figure}

Finally, in Fig.~\ref{fig:lifetime}(a) we show the spin-flip
lifetime $\tau_A=1/\Gamma$ [see Eq.~\eqref{eq:lifetime}] as a
function of temperature for an atom-surface distance of 10 $\mu$m.
The dashed line indicates the results obtained from the two-fluid
model of Ref.~\cite{skagerstam:06}. The lifetimes obtained from
Eliashberg theory (symbols) are smaller, but only by a factor of ten
or less: the influence of elastic scattering rates $\gamma$ on the
spin-flip lifetime is not very strong. This indicates that the
quality of the niobium is not critical. Surprisingly, we find that
$\tau_A$ is smallest for the high-quality sample with $\gamma=1$
meV, highest for the intermediate value $\gamma=10$ meV, and falls
off again slightly for $\gamma=100$ meV. In
Fig.~\ref{fig:lifetime}(b) we show $\tau_A$ as a function of
atom-surface distance at 4.2 K (T/T$_c=0.46$).

For an atom-surface distance of 1 $\mu$m we obtain for $\gamma=1$
meV a lifetime $\tau_A\approx 5000$ seconds at a transition
frequency of $\omega_A/2\pi=500\,\mbox{kHz}$. Values for other
distances can be obtained directly from the $z^4$ scaling of our
central equation, Eq.~\eqref{eq:lifetime}. For other (low)
frequencies, the lifetime given by Eq. \eqref{eq:lifetime} scales
approximately as $\omega_A^2$. This follows from the frequency
independence of $\sigma'$ for $\omega>0$ and the $1/\omega$
dependence of $\sigma''$.

\section{Summary}

In this article, we have resolved the controversy surrounding the
appropriate use of model assumptions for the electromagnetic energy
dissipation in superconducting materials. We have discussed the
three most common models of superconductivity, the two-fluid model,
BCS and Eliashberg theories, in ascending order of sophistication.
The spin flip lifetime of neutral atoms trapped near a
superconducting niobium surface is predicted to be much shorter when
treated in the BCS theory than it is in the two fluid model.
However,  Eliashberg theory, which improves upon the BCS theory by
including the finite quasiparticle lifetime, predicts only slightly
shorter lifetimes.

The Eliashberg theory interpolates between the two-fluid model and
the BCS theory. For intermediate scattering rates, corresponding to
real samples, the simple two-fluid model gives remarkably accurate
estimates of the Eliashberg results. We have found that the lifetime
depends only little on the precise value of the impurity scattering
rate $\gamma$.

Our numerical results based on the Eliashberg theory show that the
expected spin-flip lifetime for an atom placed one micrometer away
from a $4.2$~K superconducting planar niobium surface exceeds
several thousand seconds at an atomic transition frequency of
$500$~kHz. This is expected to scale roughly as $\omega_A^2$ and
$z^4$. Hence, superconducting surfaces provide an extremely
low-noise environment for magnetically trapped neutral atoms and
thus have great potential for coherent manipulation of atoms.

\acknowledgments

U.H. thanks Per Kristian Rekdal for helpful discussions. This work
has been supported in part by the Austrian Science Fund FWF under
project P18136--N13, by the UK Engineering and Physical Sciences
Research Council (EPSRC) Basic Technology, QIPIRC and CCM Programme
Grants, by the European Atom Chips network and by the Royal Society.

\begin{appendix}

\section{Spin flip lifetime}\label{app:lifetime}

In this appendix we sketch briefly how to derive our basic
expression \eqref{eq:lifetime}. The derivation follows closely the
general framework of Refs.~\cite{henkel:99,rekdal:04}. There is,
however, a subtle point regarding the fluctuation-dissipation
theorem, which we shall partly rephrase in the language of
solid-state physics. In the general framework developed by Welsch
and coworkers \cite{scheel:98,vogel:06,raabe:07} one introduces
Langevin noise operators with bosonic commutation relations,
in order to fulfil the linear fluctuation-dissipation
theorem. However, for calculating expectation values of bilinear
operator products as in case of the spin-flip lifetime, there is no
particular need for such an approach.

We consider an atom located in the vicinity of a dielectric body, as
depicted in Fig.~\ref{fig:schematics}, which is in a given magnetic
sublevel. The coupling to the magnetic-field fluctuations is
described through a Zeeman interaction Hamiltonian in the rotating
wave approximation. For a low-field seeking atom, the spin flip
transition is associated with an emission process, and the
transition rate is simply given by Fermi's golden rule
\cite{henkel:99,fortagh:07}
\begin{equation}\label{eq:goldenrule}
  \Gamma=\sum\limits_{\alpha,\beta}
  \frac{\langle i|\mu_\alpha|f\rangle\langle f|\mu_\beta|i\rangle}{\hbar^2}
  \,\left<B_\alpha^{\phantom\dagger}(\mathbf{r}_A,\omega_A)
  B_\beta^\dagger(\mathbf{r}_A,\omega_A)\right>\,.
\end{equation}
Here $\alpha$ and $\beta$ denote the Cartesian components, $i$ and
$f$ are the initial and final state of the scattering process,
respectively, $\mathbf{\mu}$ is the magnetic moment operator, and
$\mathbf{B}(\mathbf{r}_A,\omega)$ the Fourier transform of the
magnetic field component with positive frequency. The position of
the atom is $\mathbf{r}_A$ and $\omega_A$ is the transition
frequency.

Now we relate the spectral density of the magnetic field to the
current noise in the dielectric. In linear response theory we can
use the Green tensor $\bm G$ of Maxwell's theory to relate the
current $\mathbf{j}$ to the magnetic field $\mathbf{B}$ according to
\cite{vogel:06,rekdal:04}
\begin{equation}
  \mathbf{B}(\mathbf{r},\omega)=\mu_0 \int d^3r'\,\bm{\nabla}\times
  \bm{G}(\mathbf{r},\mathbf{r}',\omega)\,\mathbf{j}(\mathbf{r}',\omega)\,,
\end{equation}
with a corresponding equation for
$\mathbf{B}^\dagger(\mathbf{r},\omega)$. Thus, the spectral density
of the magnetic-field fluctuations is given by convolving the
spectral density $\langle
j_\alpha(\mathbf{r},\omega)j_\beta^\dagger(\mathbf{r}',\omega)\rangle$
of the current fluctuations with Maxwell's Green tensors, which
describe how the field produced by the current fluctuation
propagates to the position $\mathbf{r}_A$ of the atom (see
Fig.~\ref{fig:schematics}). In the following we consider for
simplicity only isotropic and local dielectric media.

The calculation of $\langle j(\omega)j^\dagger(\omega)\rangle$ is a
common problem in solid state physics \cite{mahan:81}. For instance,
in our present approach $j$ could be the normal current $j_n$ or the
super current $j_s$ of the superconductor. To express the spectral
density of current correlations in terms of the optical
conductivity, we first note that $\sigma'(\omega)$ is related to the
retarded current-current correlation via $\omega\sigma'(\omega)=\Re
e\int_0^\infty dt\,e^{i\omega t} \left<[j(t),j^\dagger(0)]\right>$,
which is a general result of linear-response theory \cite{kubo:85}.
A common link between the ordered and retarded current-current
correlation is provided by the spectral function
$\rho(t)=\left<[j(t),j^\dagger(0)]\right>$. Its Fourier transform
can be obtained upon insertion of a complete set of states
$|m\rangle$ with energy $E_m$ \cite{mahan:81}
\begin{eqnarray}\label{eq:spectraldensity}
  \rho(\omega)&=&\left(1-e^{-\beta\hbar\omega}\right)Z^{-1}\sum_{m,n}e^{-\beta E_m}
  \left|\langle m|j|n\rangle\right|^2\nonumber\\
  &&\qquad\times 2\pi\delta(\omega-(E_n-E_m)/\hbar)\,,
\end{eqnarray}
with $Z$ being the partition function. From Eq.~\eqref{eq:spectraldensity} one immediately obtains $\langle j(\omega)j^\dagger(\omega)\rangle=\rho(\omega)/(1-e^{-\beta\hbar\omega})$. The relation between the retarded current-current correlation and the spectral density is given through the {\em Lehmann representation}\/ as $\rho(\omega)=2\omega\sigma'(\omega)$ \cite{mahan:81}. Thus, the desired relation between $\langle j(\omega)j^\dagger(\omega)\rangle$ and the optical conductivity reads
\begin{equation}
  \langle j(\omega)j^\dagger(\omega)\rangle=
  \bigl[\bar n_{\rm th}(\hbar\omega)+1\bigr]\,2\omega\,\sigma'(\omega)\,.
\end{equation}
One finally uses the expression $\Im
m\,\bm{G}=(\mu_0/\omega)\,\bm{G}\sigma'\bm{G}^*$, which follows
directly from Maxwell's equations \cite{henry:96,scheel:98}, to
relate the scattering rate \eqref{eq:goldenrule} to the imaginary
part of the Green tensor. The final equation~\eqref{eq:lifetime} is
obtained according to the prescription given in
Ref.~\cite{skagerstam:06}.

\end{appendix}


\begin{thebibliography}{59}
\expandafter\ifx\csname natexlab\endcsname\relax\def\natexlab#1{#1}\fi
\expandafter\ifx\csname bibnamefont\endcsname\relax
  \def\bibnamefont#1{#1}\fi
\expandafter\ifx\csname bibfnamefont\endcsname\relax
  \def\bibfnamefont#1{#1}\fi
\expandafter\ifx\csname citenamefont\endcsname\relax
  \def\citenamefont#1{#1}\fi
\expandafter\ifx\csname url\endcsname\relax
  \def\url#1{\texttt{#1}}\fi
\expandafter\ifx\csname urlprefix\endcsname\relax\def\urlprefix{URL }\fi
\providecommand{\bibinfo}[2]{#2}
\providecommand{\eprint}[2][]{\url{#2}}

\bibitem[{\citenamefont{Hinds and Hughes}(1999)}]{hinds:99}
\bibinfo{author}{\bibfnamefont{E.~A.} \bibnamefont{Hinds}} \bibnamefont{and}
  \bibinfo{author}{\bibfnamefont{I.~A.} \bibnamefont{Hughes}},
  \bibinfo{journal}{J. Phys. D} \textbf{\bibinfo{volume}{32}},
  \bibinfo{pages}{R119} (\bibinfo{year}{1999}).

\bibitem[{\citenamefont{Folman et~al.}(2002)\citenamefont{Folman, Kr\"uger,
  Schmiedmayer, Denschlag, and Henkel}}]{folman:02}
\bibinfo{author}{\bibfnamefont{R.}~\bibnamefont{Folman}},
  \bibinfo{author}{\bibfnamefont{P.}~\bibnamefont{Kr\"uger}},
  \bibinfo{author}{\bibfnamefont{J.}~\bibnamefont{Schmiedmayer}},
  \bibinfo{author}{\bibfnamefont{J.}~\bibnamefont{Denschlag}},
  \bibnamefont{and} \bibinfo{author}{\bibfnamefont{C.}~\bibnamefont{Henkel}},
  \bibinfo{journal}{Adv. in Atom. Mol. and Opt. Phys.}
  \textbf{\bibinfo{volume}{48}}, \bibinfo{pages}{263} (\bibinfo{year}{2002}).

\bibitem[{spe()}]{specialissue}
\bibinfo{note}{C. Henkel, J. Schmiedmayer, and C. Westbrook, Euro. Phys. J. D
  \textbf{35}, 1 (2006) and following articles.}

\bibitem[{\citenamefont{Fortagh and Zimmermann}(2007)}]{fortagh:07}
\bibinfo{author}{\bibfnamefont{J.}~\bibnamefont{Fortagh}} \bibnamefont{and}
  \bibinfo{author}{\bibfnamefont{C.}~\bibnamefont{Zimmermann}},
  \bibinfo{journal}{Rev. Mod. Phys.} \textbf{\bibinfo{volume}{79}},
  \bibinfo{pages}{235} (\bibinfo{year}{2007}).

\bibitem[{\citenamefont{H\"ansel et~al.}(2001)\citenamefont{H\"ansel, Reichel,
  Hommelhoff, and H{\"a}nsch}}]{haensel:01}
\bibinfo{author}{\bibfnamefont{W.}~\bibnamefont{H\"ansel}},
  \bibinfo{author}{\bibfnamefont{J.}~\bibnamefont{Reichel}},
  \bibinfo{author}{\bibfnamefont{P.}~\bibnamefont{Hommelhoff}},
  \bibnamefont{and} \bibinfo{author}{\bibfnamefont{T.~W.}
  \bibnamefont{H{\"a}nsch}}, \bibinfo{journal}{Phys. Rev. A}
  \textbf{\bibinfo{volume}{64}}, \bibinfo{pages}{063607}
  (\bibinfo{year}{2001}).

\bibitem[{\citenamefont{Hinds et~al.}(2001)\citenamefont{Hinds, Vale, and
  Boshier}}]{hinds:01}
\bibinfo{author}{\bibfnamefont{E.~A.} \bibnamefont{Hinds}},
  \bibinfo{author}{\bibfnamefont{C.~J.} \bibnamefont{Vale}}, \bibnamefont{and}
  \bibinfo{author}{\bibfnamefont{M.~G.} \bibnamefont{Boshier}},
  \bibinfo{journal}{Phys. Rev. Lett.} \textbf{\bibinfo{volume}{86}},
  \bibinfo{pages}{1462} (\bibinfo{year}{2001}).

\bibitem[{\citenamefont{Andersson et~al.}(2002)\citenamefont{Andersson,
  Calarco, Folman, Andersson, Hessmo, and Schmiedmayer}}]{andersson:02}
\bibinfo{author}{\bibfnamefont{E.}~\bibnamefont{Andersson}},
  \bibinfo{author}{\bibfnamefont{T.}~\bibnamefont{Calarco}},
  \bibinfo{author}{\bibfnamefont{R.}~\bibnamefont{Folman}},
  \bibinfo{author}{\bibfnamefont{M.}~\bibnamefont{Andersson}},
  \bibinfo{author}{\bibfnamefont{B.}~\bibnamefont{Hessmo}}, \bibnamefont{and}
  \bibinfo{author}{\bibfnamefont{J.}~\bibnamefont{Schmiedmayer}},
  \bibinfo{journal}{Phys. Rev. Lett.} \textbf{\bibinfo{volume}{88}},
  \bibinfo{pages}{100401} (\bibinfo{year}{2002}).

\bibitem[{\citenamefont{Wang et~al.}(2005)\citenamefont{Wang, Anderson, Bright,
  Cornell, Diot, Kishimoto, Prentiss, Saravanan, Segal, and Wu}}]{wang:05}
\bibinfo{author}{\bibfnamefont{Y.-J.} \bibnamefont{Wang}},
  \bibinfo{author}{\bibfnamefont{D.~Z.} \bibnamefont{Anderson}},
  \bibinfo{author}{\bibfnamefont{V.~M.} \bibnamefont{Bright}},
  \bibinfo{author}{\bibfnamefont{E.~A.} \bibnamefont{Cornell}},
  \bibinfo{author}{\bibfnamefont{Q.}~\bibnamefont{Diot}},
  \bibinfo{author}{\bibfnamefont{T.}~\bibnamefont{Kishimoto}},
  \bibinfo{author}{\bibfnamefont{M.}~\bibnamefont{Prentiss}},
  \bibinfo{author}{\bibfnamefont{R.~A.} \bibnamefont{Saravanan}},
  \bibinfo{author}{\bibfnamefont{S.~R.} \bibnamefont{Segal}}, \bibnamefont{and}
  \bibinfo{author}{\bibfnamefont{S.}~\bibnamefont{Wu}}, \bibinfo{journal}{Phys.
  Rev. Lett.} \textbf{\bibinfo{volume}{94}}, \bibinfo{pages}{090405}
  (\bibinfo{year}{2005}).

\bibitem[{\citenamefont{Schumm et~al.}(2005)\citenamefont{Schumm, Hofferberth,
  Andersson, Wildermuth, Groth, Bar-Joseph, Schmiedmayer, and
  Kr\"uger}}]{schumm:05}
\bibinfo{author}{\bibfnamefont{T.}~\bibnamefont{Schumm}},
  \bibinfo{author}{\bibfnamefont{S.}~\bibnamefont{Hofferberth}},
  \bibinfo{author}{\bibfnamefont{L.~M.} \bibnamefont{Andersson}},
  \bibinfo{author}{\bibfnamefont{S.}~\bibnamefont{Wildermuth}},
  \bibinfo{author}{\bibfnamefont{S.}~\bibnamefont{Groth}},
  \bibinfo{author}{\bibfnamefont{I.}~\bibnamefont{Bar-Joseph}},
  \bibinfo{author}{\bibfnamefont{J.}~\bibnamefont{Schmiedmayer}},
  \bibnamefont{and} \bibinfo{author}{\bibfnamefont{P.}~\bibnamefont{Kr\"uger}},
  \bibinfo{journal}{Nature Phys.} \textbf{\bibinfo{volume}{1}},
  \bibinfo{pages}{57} (\bibinfo{year}{2005}).

\bibitem[{\citenamefont{Jo et~al.}(2007)\citenamefont{Jo, Shin, Will, Pasquini,
  Saba, Ketterle, Pritchard, Vengalattore, and Prentiss}}]{jo:07}
\bibinfo{author}{\bibfnamefont{G.-B.} \bibnamefont{Jo}},
  \bibinfo{author}{\bibfnamefont{Y.}~\bibnamefont{Shin}},
  \bibinfo{author}{\bibfnamefont{S.}~\bibnamefont{Will}},
  \bibinfo{author}{\bibfnamefont{T.~A.} \bibnamefont{Pasquini}},
  \bibinfo{author}{\bibfnamefont{M.}~\bibnamefont{Saba}},
  \bibinfo{author}{\bibfnamefont{W.}~\bibnamefont{Ketterle}},
  \bibinfo{author}{\bibfnamefont{D.~E.} \bibnamefont{Pritchard}},
  \bibinfo{author}{\bibfnamefont{M.}~\bibnamefont{Vengalattore}},
  \bibnamefont{and} \bibinfo{author}{\bibfnamefont{M.}~\bibnamefont{Prentiss}},
  \bibinfo{journal}{Phys. Rev. Lett.} \textbf{\bibinfo{volume}{98}},
  \bibinfo{pages}{030407} (\bibinfo{year}{2007}).

\bibitem[{\citenamefont{Calarco et~al.}(2000)\citenamefont{Calarco, Hinds,
  Jaksch, Schmiedmayer, Cirac, and Zoller}}]{calarco:00}
\bibinfo{author}{\bibfnamefont{T.}~\bibnamefont{Calarco}},
  \bibinfo{author}{\bibfnamefont{E.~A.} \bibnamefont{Hinds}},
  \bibinfo{author}{\bibfnamefont{D.}~\bibnamefont{Jaksch}},
  \bibinfo{author}{\bibfnamefont{J.}~\bibnamefont{Schmiedmayer}},
  \bibinfo{author}{\bibfnamefont{J.~I.} \bibnamefont{Cirac}}, \bibnamefont{and}
  \bibinfo{author}{\bibfnamefont{P.}~\bibnamefont{Zoller}},
  \bibinfo{journal}{Phys. Rev. A} \textbf{\bibinfo{volume}{61}},
  \bibinfo{pages}{022304} (\bibinfo{year}{2000}).

\bibitem[{\citenamefont{Charron et~al.}(2006)\citenamefont{Charron, Cirone,
  Negretti, Schmiedmayer, and Calarco}}]{charron:06}
\bibinfo{author}{\bibfnamefont{E.}~\bibnamefont{Charron}},
  \bibinfo{author}{\bibfnamefont{M.}~\bibnamefont{Cirone}},
  \bibinfo{author}{\bibfnamefont{A.}~\bibnamefont{Negretti}},
  \bibinfo{author}{\bibfnamefont{J.}~\bibnamefont{Schmiedmayer}},
  \bibnamefont{and} \bibinfo{author}{\bibfnamefont{T.}~\bibnamefont{Calarco}},
  \bibinfo{journal}{Phys. Rev. A} \textbf{\bibinfo{volume}{74}},
  \bibinfo{pages}{012308} (\bibinfo{year}{2006}).

\bibitem[{\citenamefont{Treutlein et~al.}(2006)\citenamefont{Treutlein,
  H{\"a}nsch, Reichel, Negretti, Cirone, and Calarco}}]{treutlein:06}
\bibinfo{author}{\bibfnamefont{P.}~\bibnamefont{Treutlein}},
  \bibinfo{author}{\bibfnamefont{T.~W.} \bibnamefont{H{\"a}nsch}},
  \bibinfo{author}{\bibfnamefont{J.}~\bibnamefont{Reichel}},
  \bibinfo{author}{\bibfnamefont{A.}~\bibnamefont{Negretti}},
  \bibinfo{author}{\bibfnamefont{M.~A.} \bibnamefont{Cirone}},
  \bibnamefont{and} \bibinfo{author}{\bibfnamefont{T.}~\bibnamefont{Calarco}},
  \bibinfo{journal}{Phys. Rev. A} \textbf{\bibinfo{volume}{74}},
  \bibinfo{pages}{022312} (\bibinfo{year}{2006}).

\bibitem[{\citenamefont{Hohenester et~al.}(2007)\citenamefont{Hohenester,
  Rekdal, Borzi, and Schmiedmayer}}]{hohenester.pra:07}
\bibinfo{author}{\bibfnamefont{U.}~\bibnamefont{Hohenester}},
  \bibinfo{author}{\bibfnamefont{P.~K.} \bibnamefont{Rekdal}},
  \bibinfo{author}{\bibfnamefont{A.}~\bibnamefont{Borzi}}, \bibnamefont{and}
  \bibinfo{author}{\bibfnamefont{J.}~\bibnamefont{Schmiedmayer}},
  \bibinfo{journal}{Phys. Rev. A} \textbf{\bibinfo{volume}{75}},
  \bibinfo{pages}{023602} (\bibinfo{year}{2007}).

\bibitem[{\citenamefont{Paul et~al.}(2005)\citenamefont{Paul, Richter, and
  Schlagheck}}]{paul:05}
\bibinfo{author}{\bibfnamefont{T.}~\bibnamefont{Paul}},
  \bibinfo{author}{\bibfnamefont{K.}~\bibnamefont{Richter}}, \bibnamefont{and}
  \bibinfo{author}{\bibfnamefont{P.}~\bibnamefont{Schlagheck}},
  \bibinfo{journal}{Phys. Rev. Lett.} \textbf{\bibinfo{volume}{94}},
  \bibinfo{pages}{020404} (\bibinfo{year}{2005}).

\bibitem[{\citenamefont{Wildermuth et~al.}(2005)\citenamefont{Wildermuth,
  Hofferberth, Lesanovsky, Haller, Mauritz-Andersson, Groth, Bar-Joseph,
  Kr{\"u}ger, and Schmiedmayer}}]{wildermuth:05}
\bibinfo{author}{\bibfnamefont{S.}~\bibnamefont{Wildermuth}},
  \bibinfo{author}{\bibfnamefont{S.}~\bibnamefont{Hofferberth}},
  \bibinfo{author}{\bibfnamefont{I.}~\bibnamefont{Lesanovsky}},
  \bibinfo{author}{\bibfnamefont{E.}~\bibnamefont{Haller}},
  \bibinfo{author}{\bibfnamefont{L.}~\bibnamefont{Mauritz-Andersson}},
  \bibinfo{author}{\bibfnamefont{S.}~\bibnamefont{Groth}},
  \bibinfo{author}{\bibfnamefont{I.}~\bibnamefont{Bar-Joseph}},
  \bibinfo{author}{\bibfnamefont{P.}~\bibnamefont{Kr{\"u}ger}},
  \bibnamefont{and}
  \bibinfo{author}{\bibfnamefont{J.}~\bibnamefont{Schmiedmayer}},
  \bibinfo{journal}{Nature (London)} \textbf{\bibinfo{volume}{435}},
  \bibinfo{pages}{440} (\bibinfo{year}{2005}).

\bibitem[{\citenamefont{Scheel et~al.}(2007)\citenamefont{Scheel, Fermani, and
  Hinds}}]{scheel:07}
\bibinfo{author}{\bibfnamefont{S.}~\bibnamefont{Scheel}},
  \bibinfo{author}{\bibfnamefont{R.}~\bibnamefont{Fermani}}, \bibnamefont{and}
  \bibinfo{author}{\bibfnamefont{E.~A.} \bibnamefont{Hinds}},
  \bibinfo{journal}{Phys. Rev. A} \textbf{\bibinfo{volume}{75}},
  \bibinfo{pages}{064901} (\bibinfo{year}{2007}).

\bibitem[{\citenamefont{Jones et~al.}(2003)\citenamefont{Jones, Vale, Sahagun,
  Hall, and Hinds}}]{jones:03}
\bibinfo{author}{\bibfnamefont{M.~P.~A.} \bibnamefont{Jones}},
  \bibinfo{author}{\bibfnamefont{C.~J.} \bibnamefont{Vale}},
  \bibinfo{author}{\bibfnamefont{D.}~\bibnamefont{Sahagun}},
  \bibinfo{author}{\bibfnamefont{B.~V.} \bibnamefont{Hall}}, \bibnamefont{and}
  \bibinfo{author}{\bibfnamefont{E.~A.} \bibnamefont{Hinds}},
  \bibinfo{journal}{Phys. Rev. Lett.} \textbf{\bibinfo{volume}{91}},
  \bibinfo{pages}{080401} (\bibinfo{year}{2003}).

\bibitem[{\citenamefont{Lin et~al.}(2004)\citenamefont{Lin, Teper, Chin, and
  Vuletic}}]{lin:04}
\bibinfo{author}{\bibfnamefont{Y.}~\bibnamefont{Lin}},
  \bibinfo{author}{\bibfnamefont{I.}~\bibnamefont{Teper}},
  \bibinfo{author}{\bibfnamefont{C.}~\bibnamefont{Chin}}, \bibnamefont{and}
  \bibinfo{author}{\bibfnamefont{V.}~\bibnamefont{Vuletic}},
  \bibinfo{journal}{Phys. Rev. Lett.} \textbf{\bibinfo{volume}{92}},
  \bibinfo{pages}{050404} (\bibinfo{year}{2004}).

\bibitem[{\citenamefont{Henkel et~al.}(1999)\citenamefont{Henkel, P\"otting,
  and Wilkens}}]{henkel:99}
\bibinfo{author}{\bibfnamefont{C.}~\bibnamefont{Henkel}},
  \bibinfo{author}{\bibfnamefont{S.}~\bibnamefont{P\"otting}},
  \bibnamefont{and} \bibinfo{author}{\bibfnamefont{M.}~\bibnamefont{Wilkens}},
  \bibinfo{journal}{Appl. Phys. B} \textbf{\bibinfo{volume}{69}},
  \bibinfo{pages}{379} (\bibinfo{year}{1999}).

\bibitem[{\citenamefont{Rekdal et~al.}(2004)\citenamefont{Rekdal, Scheel,
  Knight, and Hinds}}]{rekdal:04}
\bibinfo{author}{\bibfnamefont{P.~K.} \bibnamefont{Rekdal}},
  \bibinfo{author}{\bibfnamefont{S.}~\bibnamefont{Scheel}},
  \bibinfo{author}{\bibfnamefont{P.~L.} \bibnamefont{Knight}},
  \bibnamefont{and} \bibinfo{author}{\bibfnamefont{E.~A.} \bibnamefont{Hinds}},
  \bibinfo{journal}{Phys. Rev. A} \textbf{\bibinfo{volume}{70}},
  \bibinfo{pages}{013811} (\bibinfo{year}{2004}).

\bibitem[{\citenamefont{Scheel et~al.}(2005)\citenamefont{Scheel, Rekdal,
  Knight, and Hinds}}]{scheel:05}
\bibinfo{author}{\bibfnamefont{S.}~\bibnamefont{Scheel}},
  \bibinfo{author}{\bibfnamefont{P.~K.} \bibnamefont{Rekdal}},
  \bibinfo{author}{\bibfnamefont{P.~L.} \bibnamefont{Knight}},
  \bibnamefont{and} \bibinfo{author}{\bibfnamefont{E.~A.} \bibnamefont{Hinds}},
  \bibinfo{journal}{Phys. Rev. A} \textbf{\bibinfo{volume}{72}},
  \bibinfo{pages}{042901} (\bibinfo{year}{2005}).

\bibitem[{\citenamefont{Nirrengarten et~al.}(2006)\citenamefont{Nirrengarten,
  Qarry, Roux, Emmert, Nogues, Brune, Raimond, and Haroche}}]{nirrengarten:06}
\bibinfo{author}{\bibfnamefont{T.}~\bibnamefont{Nirrengarten}},
  \bibinfo{author}{\bibfnamefont{A.}~\bibnamefont{Qarry}},
  \bibinfo{author}{\bibfnamefont{C.}~\bibnamefont{Roux}},
  \bibinfo{author}{\bibfnamefont{A.}~\bibnamefont{Emmert}},
  \bibinfo{author}{\bibfnamefont{G.}~\bibnamefont{Nogues}},
  \bibinfo{author}{\bibfnamefont{M.}~\bibnamefont{Brune}},
  \bibinfo{author}{\bibfnamefont{J.~M.} \bibnamefont{Raimond}},
  \bibnamefont{and} \bibinfo{author}{\bibfnamefont{S.}~\bibnamefont{Haroche}},
  \bibinfo{journal}{Phys. Rev. Lett.} \textbf{\bibinfo{volume}{97}},
  \bibinfo{pages}{200405} (\bibinfo{year}{2006}).

\bibitem[{\citenamefont{Mukai et~al.}(2007)\citenamefont{Mukai, Hufnagel,
  Kasper, Meno, Tsukada, Semba, and Shimizu}}]{shimizu:07}
\bibinfo{author}{\bibfnamefont{T.}~\bibnamefont{Mukai}},
  \bibinfo{author}{\bibfnamefont{C.}~\bibnamefont{Hufnagel}},
  \bibinfo{author}{\bibfnamefont{A.}~\bibnamefont{Kasper}},
  \bibinfo{author}{\bibfnamefont{T.}~\bibnamefont{Meno}},
  \bibinfo{author}{\bibfnamefont{A.}~\bibnamefont{Tsukada}},
  \bibinfo{author}{\bibfnamefont{K.}~\bibnamefont{Semba}}, \bibnamefont{and}
  \bibinfo{author}{\bibfnamefont{F.}~\bibnamefont{Shimizu}},
  \bibinfo{journal}{Phys. Rev. Lett.} \textbf{\bibinfo{volume}{98}},
  \bibinfo{pages}{260407} (\bibinfo{year}{2007}).

\bibitem[{\citenamefont{Skagerstam et~al.}(2006)\citenamefont{Skagerstam,
  Hohenester, Eiguren, and Rekdal}}]{skagerstam:06}
\bibinfo{author}{\bibfnamefont{B.~S.} \bibnamefont{Skagerstam}},
  \bibinfo{author}{\bibfnamefont{U.}~\bibnamefont{Hohenester}},
  \bibinfo{author}{\bibfnamefont{A.}~\bibnamefont{Eiguren}}, \bibnamefont{and}
  \bibinfo{author}{\bibfnamefont{P.~K.} \bibnamefont{Rekdal}},
  \bibinfo{journal}{Phys. Rev. Lett.} \textbf{\bibinfo{volume}{97}},
  \bibinfo{pages}{070401} (\bibinfo{year}{2006}).

\bibitem[{\citenamefont{Scheel et~al.}()\citenamefont{Scheel, Hinds, and
  Knight}}]{scheel.comment:06}
\bibinfo{author}{\bibfnamefont{S.}~\bibnamefont{Scheel}},
  \bibinfo{author}{\bibfnamefont{E.~A.} \bibnamefont{Hinds}}, \bibnamefont{and}
  \bibinfo{author}{\bibfnamefont{P.~L.} \bibnamefont{Knight}},
  \bibinfo{note}{arXiv:quant-ph/0610095}.

\bibitem[{\citenamefont{Skagerstam et~al.}()\citenamefont{Skagerstam,
  Hohenester, Eiguren, and Rekdal}}]{skagerstam.reply:06}
\bibinfo{author}{\bibfnamefont{B.~S.} \bibnamefont{Skagerstam}},
  \bibinfo{author}{\bibfnamefont{U.}~\bibnamefont{Hohenester}},
  \bibinfo{author}{\bibfnamefont{A.}~\bibnamefont{Eiguren}}, \bibnamefont{and}
  \bibinfo{author}{\bibfnamefont{P.~K.} \bibnamefont{Rekdal}},
  \bibinfo{note}{arXiv:quant-ph/061025}.

\bibitem[{\citenamefont{Bardeen et~al.}(1957)\citenamefont{Bardeen, Cooper, and
  Schrieffer}}]{bardeen:57}
\bibinfo{author}{\bibfnamefont{J.}~\bibnamefont{Bardeen}},
  \bibinfo{author}{\bibfnamefont{L.~N.} \bibnamefont{Cooper}},
  \bibnamefont{and} \bibinfo{author}{\bibfnamefont{J.~R.}
  \bibnamefont{Schrieffer}}, \bibinfo{journal}{Phys. Rev.}
  \textbf{\bibinfo{volume}{108}}, \bibinfo{pages}{1175} (\bibinfo{year}{1957}).

\bibitem[{\citenamefont{Eliashberg}(1960)}]{eliashberg:60}
\bibinfo{author}{\bibfnamefont{G.~M.} \bibnamefont{Eliashberg}},
  \bibinfo{journal}{Soviet. Phys. JETP} \textbf{\bibinfo{volume}{11}},
  \bibinfo{pages}{696} (\bibinfo{year}{1960}).

\bibitem[{\citenamefont{Marsiglio and Carbotte}(2002)}]{marsiglio:02}
\bibinfo{author}{\bibfnamefont{F.}~\bibnamefont{Marsiglio}} \bibnamefont{and}
  \bibinfo{author}{\bibfnamefont{J.~P.} \bibnamefont{Carbotte}},
  \emph{\bibinfo{title}{The Physics of Superconductors}}
  (\bibinfo{publisher}{Springer, Berlin}, \bibinfo{year}{2002}), p.
  \bibinfo{pages}{233}.

\bibitem[{\citenamefont{Rickayzen}(1965)}]{rickayzen:65}
\bibinfo{author}{\bibfnamefont{G.}~\bibnamefont{Rickayzen}},
  \emph{\bibinfo{title}{Theory of Superconductivity}}
  (\bibinfo{publisher}{Interscience, New York}, \bibinfo{year}{1965}).

\bibitem[{\citenamefont{Tinkham}(1975)}]{tinkham:75}
\bibinfo{author}{\bibfnamefont{M.}~\bibnamefont{Tinkham}},
  \emph{\bibinfo{title}{Introduction to Superconductivity}}
  (\bibinfo{publisher}{McGraw-Hill, New York}, \bibinfo{year}{1975}).

\bibitem[{\citenamefont{Mahan}(1981)}]{mahan:81}
\bibinfo{author}{\bibfnamefont{G.~D.} \bibnamefont{Mahan}},
  \emph{\bibinfo{title}{Many-Particle Physics}} (\bibinfo{publisher}{Plenum},
  \bibinfo{address}{New York}, \bibinfo{year}{1981}).

\bibitem[{\citenamefont{London and London}(1935)}]{london:35}
\bibinfo{author}{\bibfnamefont{F.}~\bibnamefont{London}} \bibnamefont{and}
  \bibinfo{author}{\bibfnamefont{H.}~\bibnamefont{London}},
  \bibinfo{journal}{Proc. Roy. Soc. Lond.} \textbf{\bibinfo{volume}{133}},
  \bibinfo{pages}{497} (\bibinfo{year}{1935}).

\bibitem[{\citenamefont{Gorter and Casimir}(1934)}]{gorter:34}
\bibinfo{author}{\bibfnamefont{C.~S.} \bibnamefont{Gorter}} \bibnamefont{and}
  \bibinfo{author}{\bibfnamefont{H.}~\bibnamefont{Casimir}},
  \bibinfo{journal}{Z. Phys.} \textbf{\bibinfo{volume}{35}},
  \bibinfo{pages}{963} (\bibinfo{year}{1934}).

\bibitem[{\citenamefont{Hebel and Slichter}(1959)}]{hebel:59}
\bibinfo{author}{\bibfnamefont{L.~C.} \bibnamefont{Hebel}} \bibnamefont{and}
  \bibinfo{author}{\bibfnamefont{C.~P.} \bibnamefont{Slichter}},
  \bibinfo{journal}{Phys. Rev.} \textbf{\bibinfo{volume}{113}},
  \bibinfo{pages}{1504} (\bibinfo{year}{1959}).

\bibitem[{\citenamefont{Fr{\"o}hlich}(1950)}]{froehlich:50}
\bibinfo{author}{\bibfnamefont{H.}~\bibnamefont{Fr{\"o}hlich}},
  \bibinfo{journal}{Phys. Rev.} \textbf{\bibinfo{volume}{79}},
  \bibinfo{pages}{845} (\bibinfo{year}{1950}).

\bibitem[{\citenamefont{Cooper}(1956)}]{cooper:56}
\bibinfo{author}{\bibfnamefont{L.~N.} \bibnamefont{Cooper}},
  \bibinfo{journal}{Phys. Rev.} \textbf{\bibinfo{volume}{104}},
  \bibinfo{pages}{1189} (\bibinfo{year}{1956}).

\bibitem[{\citenamefont{Mattis and Bardeen}(1958)}]{mattis:58}
\bibinfo{author}{\bibfnamefont{D.~C.} \bibnamefont{Mattis}} \bibnamefont{and}
  \bibinfo{author}{\bibfnamefont{J.}~\bibnamefont{Bardeen}},
  \bibinfo{journal}{Phys. Rev.} \textbf{\bibinfo{volume}{111}},
  \bibinfo{pages}{412} (\bibinfo{year}{1958}).

\bibitem[{\citenamefont{Klein et~al.}(1984)\citenamefont{Klein, Nicol, Holczer,
  and Gr{\"u}ner}}]{klein:84}
\bibinfo{author}{\bibfnamefont{O.}~\bibnamefont{Klein}},
  \bibinfo{author}{\bibfnamefont{E.~J.} \bibnamefont{Nicol}},
  \bibinfo{author}{\bibfnamefont{K.}~\bibnamefont{Holczer}}, \bibnamefont{and}
  \bibinfo{author}{\bibfnamefont{G.}~\bibnamefont{Gr{\"u}ner}},
  \bibinfo{journal}{Phys. Rev. B} \textbf{\bibinfo{volume}{50}},
  \bibinfo{pages}{6307} (\bibinfo{year}{1984}).

\bibitem[{\citenamefont{Marsiglio et~al.}(1988)\citenamefont{Marsiglio,
  Schossmann, and Carbotte}}]{marsiglio:88}
\bibinfo{author}{\bibfnamefont{F.}~\bibnamefont{Marsiglio}},
  \bibinfo{author}{\bibfnamefont{M.}~\bibnamefont{Schossmann}},
  \bibnamefont{and} \bibinfo{author}{\bibfnamefont{J.~P.}
  \bibnamefont{Carbotte}}, \bibinfo{journal}{Phys. Rev. B}
  \textbf{\bibinfo{volume}{37}}, \bibinfo{pages}{4965} (\bibinfo{year}{1988}).

\bibitem[{\citenamefont{Carbotte}(1990)}]{carbotte:90}
\bibinfo{author}{\bibfnamefont{J.~P.} \bibnamefont{Carbotte}},
  \bibinfo{journal}{Rev. Mod. Phys.} \textbf{\bibinfo{volume}{62}},
  \bibinfo{pages}{1027} (\bibinfo{year}{1990}).

\bibitem[{\citenamefont{Nicol and Carbotte}(1992)}]{nicol:92}
\bibinfo{author}{\bibfnamefont{E.~J.} \bibnamefont{Nicol}} \bibnamefont{and}
  \bibinfo{author}{\bibfnamefont{J.~P.} \bibnamefont{Carbotte}},
  \bibinfo{journal}{Phys. Rev. B} \textbf{\bibinfo{volume}{45}},
  \bibinfo{pages}{10519} (\bibinfo{year}{1992}).

\bibitem[{nam()}]{nam:67}
\bibinfo{note}{S. B.~Nam, Phys. Rev. \textbf{156}, 470 (1967); \textit{ibid.}
  \textbf{156}, 487 (1967).}

\bibitem[{\citenamefont{Anderson}(1959)}]{anderson:59}
\bibinfo{author}{\bibfnamefont{P.~W.} \bibnamefont{Anderson}},
  \bibinfo{journal}{J. Phys. Chem- Solids} \textbf{\bibinfo{volume}{11}},
  \bibinfo{pages}{26} (\bibinfo{year}{1959}).

\bibitem[{\citenamefont{Balatsky et~al.}(2006)\citenamefont{Balatsky, Vekhter,
  and Zhu}}]{balatsky:06}
\bibinfo{author}{\bibfnamefont{A.~V.} \bibnamefont{Balatsky}},
  \bibinfo{author}{\bibfnamefont{I.}~\bibnamefont{Vekhter}}, \bibnamefont{and}
  \bibinfo{author}{\bibfnamefont{J.~X.} \bibnamefont{Zhu}},
  \bibinfo{journal}{Rev. Mod. Phys.} \textbf{\bibinfo{volume}{78}},
  \bibinfo{pages}{373} (\bibinfo{year}{2006}).

\bibitem[{\citenamefont{Casalbuoni et~al.}(2005)\citenamefont{Casalbuoni,
  Knabbe, K{\"o}tzler, Lilje, v{on S}awiliski, Schm{\"u}ser, and
  Steffen}}]{casalbuoni:05}
\bibinfo{author}{\bibfnamefont{S.}~\bibnamefont{Casalbuoni}},
  \bibinfo{author}{\bibfnamefont{E.~A.} \bibnamefont{Knabbe}},
  \bibinfo{author}{\bibfnamefont{J.}~\bibnamefont{K{\"o}tzler}},
  \bibinfo{author}{\bibfnamefont{L.}~\bibnamefont{Lilje}},
  \bibinfo{author}{\bibfnamefont{L.}~\bibnamefont{v{on S}awiliski}},
  \bibinfo{author}{\bibfnamefont{P.}~\bibnamefont{Schm{\"u}ser}},
  \bibnamefont{and} \bibinfo{author}{\bibfnamefont{B.}~\bibnamefont{Steffen}},
  \bibinfo{journal}{Nucl. Instrum. Meth. Phys. Res. A}
  \textbf{\bibinfo{volume}{538}}, \bibinfo{pages}{45} (\bibinfo{year}{2005}).

\bibitem[{\citenamefont{Perkowitz et~al.}(1985)\citenamefont{Perkowitz, Carr,
  Subramaniam, and Mitrovic}}]{perkowitz:85}
\bibinfo{author}{\bibfnamefont{S.}~\bibnamefont{Perkowitz}},
  \bibinfo{author}{\bibfnamefont{G.~L.} \bibnamefont{Carr}},
  \bibinfo{author}{\bibfnamefont{B.}~\bibnamefont{Subramaniam}},
  \bibnamefont{and} \bibinfo{author}{\bibfnamefont{B.}~\bibnamefont{Mitrovic}},
  \bibinfo{journal}{Phys. Rev. B} \textbf{\bibinfo{volume}{32}},
  \bibinfo{pages}{153} (\bibinfo{year}{1985}).

\bibitem[{\citenamefont{Pronin et~al.}(1998)\citenamefont{Pronin, Dressel,
  Pimenov, Loidl, Roshchin, and Greene}}]{pronin:98}
\bibinfo{author}{\bibfnamefont{A.~V.} \bibnamefont{Pronin}},
  \bibinfo{author}{\bibfnamefont{M.}~\bibnamefont{Dressel}},
  \bibinfo{author}{\bibfnamefont{A.}~\bibnamefont{Pimenov}},
  \bibinfo{author}{\bibfnamefont{A.}~\bibnamefont{Loidl}},
  \bibinfo{author}{\bibfnamefont{I.~V.} \bibnamefont{Roshchin}},
  \bibnamefont{and} \bibinfo{author}{\bibfnamefont{L.~H.}
  \bibnamefont{Greene}}, \bibinfo{journal}{Phys. Rev. B}
  \textbf{\bibinfo{volume}{57}}, \bibinfo{pages}{14416} (\bibinfo{year}{1998}).

\bibitem[{\citenamefont{Ashcroft and Mermin}(1976)}]{ashcroft:76}
\bibinfo{author}{\bibfnamefont{N.~W.} \bibnamefont{Ashcroft}} \bibnamefont{and}
  \bibinfo{author}{\bibfnamefont{N.~D.} \bibnamefont{Mermin}},
  \emph{\bibinfo{title}{Solid State Physics}} (\bibinfo{publisher}{Saunders,
  Fort Worth}, \bibinfo{year}{1976}).

\bibitem[{\citenamefont{Miller}(1959)}]{miller:59}
\bibinfo{author}{\bibfnamefont{P.~B.} \bibnamefont{Miller}},
  \bibinfo{journal}{Phys. Rev.} \textbf{\bibinfo{volume}{113}},
  \bibinfo{pages}{1209} (\bibinfo{year}{1959}).

\bibitem[{\citenamefont{Baroni et~al.}(2001)\citenamefont{Baroni, de~Gironcoli,
  Dal~Corso, and Giannozzi}}]{baroni:01}
\bibinfo{author}{\bibfnamefont{S.}~\bibnamefont{Baroni}},
  \bibinfo{author}{\bibfnamefont{S.}~\bibnamefont{de~Gironcoli}},
  \bibinfo{author}{\bibfnamefont{A.}~\bibnamefont{Dal~Corso}},
  \bibnamefont{and}
  \bibinfo{author}{\bibfnamefont{P.}~\bibnamefont{Giannozzi}},
  \bibinfo{journal}{Rev. Mod. Phys.} \textbf{\bibinfo{volume}{73}},
  \bibinfo{pages}{515} (\bibinfo{year}{2001}).

\bibitem[{\citenamefont{Savrasov and Savrasov}(1996)}]{savrasov:96}
\bibinfo{author}{\bibfnamefont{S.~Y.} \bibnamefont{Savrasov}} \bibnamefont{and}
  \bibinfo{author}{\bibfnamefont{D.~Y.} \bibnamefont{Savrasov}},
  \bibinfo{journal}{Phys. Rev. B} \textbf{\bibinfo{volume}{54}},
  \bibinfo{pages}{16487} (\bibinfo{year}{1996}).

\bibitem[{\citenamefont{Marsiglio et~al.}(1994)\citenamefont{Marsiglio,
  Carbotte, Akis, Achkir, and Poirier}}]{marsiglio:94}
\bibinfo{author}{\bibfnamefont{F.}~\bibnamefont{Marsiglio}},
  \bibinfo{author}{\bibfnamefont{J.~P.} \bibnamefont{Carbotte}},
  \bibinfo{author}{\bibfnamefont{R.}~\bibnamefont{Akis}},
  \bibinfo{author}{\bibfnamefont{D.}~\bibnamefont{Achkir}}, \bibnamefont{and}
  \bibinfo{author}{\bibfnamefont{M.}~\bibnamefont{Poirier}},
  \bibinfo{journal}{Phys. Rev. B} \textbf{\bibinfo{volume}{50}},
  \bibinfo{pages}{7203} (\bibinfo{year}{1994}).

\bibitem[{\citenamefont{Scheel et~al.}(1998)\citenamefont{Scheel, Kn\"oll, and
  Welsch}}]{scheel:98}
\bibinfo{author}{\bibfnamefont{S.}~\bibnamefont{Scheel}},
  \bibinfo{author}{\bibfnamefont{L.}~\bibnamefont{Kn\"oll}}, \bibnamefont{and}
  \bibinfo{author}{\bibfnamefont{D.-G.} \bibnamefont{Welsch}},
  \bibinfo{journal}{Phys. Rev. A} \textbf{\bibinfo{volume}{58}},
  \bibinfo{pages}{700} (\bibinfo{year}{1998}).

\bibitem[{\citenamefont{Vogel and Welsch}(2006)}]{vogel:06}
\bibinfo{author}{\bibfnamefont{W.}~\bibnamefont{Vogel}} \bibnamefont{and}
  \bibinfo{author}{\bibfnamefont{D.-G.} \bibnamefont{Welsch}},
  \emph{\bibinfo{title}{Quantum Optics}} (\bibinfo{publisher}{Wiley, Berlin},
  \bibinfo{year}{2006}).

\bibitem[{\citenamefont{Raabe et~al.}(2007)\citenamefont{Raabe, Scheel, and
  Welsch}}]{raabe:07}
\bibinfo{author}{\bibfnamefont{C.}~\bibnamefont{Raabe}},
  \bibinfo{author}{\bibfnamefont{S.}~\bibnamefont{Scheel}}, \bibnamefont{and}
  \bibinfo{author}{\bibfnamefont{D.~G.} \bibnamefont{Welsch}},
  \bibinfo{journal}{Phys. Rev. A} \textbf{\bibinfo{volume}{75}},
  \bibinfo{pages}{053813} (\bibinfo{year}{2007}).

\bibitem[{\citenamefont{Kubo et~al.}(1985)\citenamefont{Kubo, Toda, and
  Hashitsume}}]{kubo:85}
\bibinfo{author}{\bibfnamefont{R.}~\bibnamefont{Kubo}},
  \bibinfo{author}{\bibfnamefont{M.}~\bibnamefont{Toda}}, \bibnamefont{and}
  \bibinfo{author}{\bibfnamefont{M.}~\bibnamefont{Hashitsume}},
  \emph{\bibinfo{title}{Statistical Physics {II}}}
  (\bibinfo{publisher}{Springer}, \bibinfo{address}{Berlin},
  \bibinfo{year}{1985}).

\bibitem[{\citenamefont{Henry and Kazarinov}(1996)}]{henry:96}
\bibinfo{author}{\bibfnamefont{C.~H.} \bibnamefont{Henry}} \bibnamefont{and}
  \bibinfo{author}{\bibfnamefont{R.~F.} \bibnamefont{Kazarinov}},
  \bibinfo{journal}{Rev. Mod. Phys.} \textbf{\bibinfo{volume}{68}},
  \bibinfo{pages}{801} (\bibinfo{year}{1996}).

\end{thebibliography}
\end{document}